\documentclass[fleqn,usenatbib]{mnras}

\usepackage{newtxtext,newtxmath}
\usepackage[T1]{fontenc}

\DeclareRobustCommand{\VAN}[3]{#2}
\let\VANthebibliography\thebibliography
\def\thebibliography{\DeclareRobustCommand{\VAN}[3]{##3}\VANthebibliography}

\usepackage{graphicx}	% Including figure files
\usepackage{amsmath}	% Advanced maths commands
\usepackage{bm}
\usepackage[mathscr]{eucal}
\usepackage{soul}
\usepackage{xcolor}
\usepackage{booktabs}

\soulregister\citep7
\title[WD-WD collisions in AGN discs]{White Dwarf--White Dwarf collisions in AGN discs via close encounters}

\author[Yan Luo et al.]{
Yan Luo,$^{1,2}$\thanks{E-mail: dearye@mail.ustc.edu.cn(YL)}
Xiao-Jun Wu,$^{1,2}$
Shu-Rui Zhang,$^{1,2}$
Jian-Min Wang, $^{3}$
Luis C. Ho, $^{4}$
Ye-Fei Yuan$^{1,2}$\thanks{E-mail: yfyuan@ustc.edu.cn(YFY)}
\\
$^{1}$School of Astronomy and Space Science, University of Science and Technology of China, Hefei 230026, China\\
$^{2}$CAS Key Laboratory for Research in Galaxies and Cosmology, Department of Astronomy, University of Science and Technology of China, Hefei 230026, China\\
$^{3}$Key Laboratory for Particle Astrophysics, Institute of High Energy Physics, Chinese Academy of Sciences, 19B Yuquan Road, Beijing 100049, China\\
$^{4}$Kavli Institute for Astronomy and Astrophysics, Peking University, Beijing 100871, China
}

\date{Accepted 2023 July 18. Received 2023 June 24; in original form 2022 September 30}

\pubyear{2022}

\begin{document}
\label{firstpage}
\pagerange{\pageref{firstpage}--\pageref{lastpage}}
\maketitle

\begin{abstract}
White dwarfs (WDs) in active galactic nucleus (AGNs) discs might migrate to the inner radii of the discs and form restricted three-body systems with two WDs moving around the central supermassive black hole (SMBH) in close orbits. These systems could be dynamical unstable, which can lead to very close encounters or direct collisions. In this work, we use N-body simulations to study the evolution of such systems with the different initial orbital separation $p$, relative orbital inclination $\Delta{i}$ and SMBH mass $M$. It is found that the close encounters of WDs mainly occur at $1.1R_{\rm H} \lesssim p \lesssim 2\sqrt{3}R_{\rm H}$, where $R_{\rm H}$ is the mutual Hill radius. For $p<1.1R_{\rm H}$, the majority of WDs move in horseshoe or tadpole orbits, and only few of them with small initial orbital phase difference undergo close encounters. For $p=3.0R_{\rm H}$, WD-WD collisions occur in most of the samples within a time of $10^5P_1$, and considerable collisions occur within a time of $t<62P_1$ for small orbital radii, where $P_1$ is the orbital period. The peak of the closest separation distribution increase and the WD-WD collision fraction decreases with an increase of the relative inclination. The closest separation distribution is similar in cases with the different SMBH mass, but the WD-WD collision fraction decreases as the mass of SMBHs increases. According to our estimation, the event rate of the cosmic WD-WD collision in AGN discs is about $300{\rm Gpc^{-3}yr^{-1}}$, roughly $1\%$ of the one of the observed type Ia supernova. The corresponding electromagnetic emission signals can be observed by large surveys of AGNs.
\end{abstract}


%%%%%%%%%%%%%%%%%%%%%%%%%%%%%%%%%%%%%%%%%%%%%%%%%%

%%%%%%%%%%%%%%%%% BODY OF PAPER %%%%%%%%%%%%%%%%%%

\section{Introduction}
Active galactic nuclei (AGNs) are considered to be an  important astrophysics environment for mergers and collisions of binary compact objects and gravitational wave events \citep{1999ApJ...521..502C}, and have recently been the subject of renew interest
%In recent years, active galactic nuclei (AGNs) have been considered to be important astrophysics laboratories for producing compact object mergers, collisions and gravitational wave events 
\citep{2020ApJ...898...25T, 2020MNRAS.494.1203M, 2020MNRAS.498.4088M, 2022PhRvD.105f3006L, 2018ApJ...866...66M, 2022arXiv220207633L, 2017ApJ...835..165B}. Due to their deep potential and high gas density, AGNs can retain a large population of stellar remnants and lead to hierarchical merger \citep{2019PhRvL.123r1101Y, 2020ApJ...899...26T,  2021arXiv211010838W, 2021MNRAS.502.2049L}. These compact objects are probably derived from nuclear star clusters or stellar evolution in situ in the extended region ($\sim$ pc) of the AGN discs \citep{2017MNRAS.464..946S, 2020MNRAS.493.3732D}. In this environment, compact objects may be captured and align with the AGN discs. \citep{2019ApJ...876..122Y}. \citet{Tanaka_2002, 2010MNRAS.401.1950P, 2012MNRAS.425..460M} indicate that the torque exerted by the disc 
helps them move to the migration trap \citep{2016ApJ...819L..17B, Secunda_2019, 2020ApJ...903..133S} within the disc. Normally, the migration traps are located around $20-300$ Schwarzschild radius. Thus, It is not unreasonable to expect that many of these compact objects will be assembled in the AGN discs, which could be responsible for some LIGO/Virgo black hole (BH) binary merger events. Due to high gas density, these compact objects accrete gas in AGN discs \citep{2021ApJ...923..173P, 2021ApJ...911L..14W}, and mergers of these compact objects may potentially be associated with some electromagnetic radiation \citep{1999ApJ...521..502C, 2021ApJ...916L..17W}.

Following the LIGO/Virgo collaborations \citep{2019ApJ...882L..24A, 2021ApJ...913L...7A,  2021arXiv210801045T}, BH-BH mergers in AGN discs have attracted much attention \citep{2021ApJ...920L..42G, 2020arXiv201009765S, 2020MNRAS.499.2608F, 2021ApJ...907L..20T, 2020A&A...638A.119G}. In the AGN discs, the enrichment of compact objects enhances the mergers between compact objects. The initial mass function (IMF) \citep{2001MNRAS.322..231K} suggests that there are more neutron stars (NSs) and white dwarfs (WDs) than BHs exist in the AGN discs. These high-density WDs within the disc might lead to tidal disruption of WD by BH \citep{2012ApJ...749..117H, 2018ApJ...865....3A, 2018MNRAS.477.3449K, 2018ApJ...867..119F} and supernovae due to the merger of  WDs \citep{2018ApJ...869..140K, 2013ApJ...773..136J}. As a result, the associated electromagnetic radiation of such events might be detected by large surveys of AGNs \citep{2017MNRAS.470.4112G, 2020MNRAS.493..477C}.

Once a compact objects align with the disc, it can form a restricted three-body systems with two compact objects moving around the  central supermassive black hole (SMBH) in closely packed circular orbits. The orbits of these restricted three-body systems may be unstable, leading to close encounters and Jacobi captures \citep{2002Natur.420..643G, 2021MNRAS.506.1665G, 2022arXiv220309646B}. 
%These close encounters and Jacobi captures between two compact objects may be such a mechanism. 
When the separation of two compact objects is smaller than a critical radius, known as the Hill radius or Jacobi radius, where the gravitation force between two compact objects is comparable to the tidal force of the central SMBH, the evolution of system turns to chaos and close encounters and Jacobi captures become possible. During the close encounters, the two compact objects may dissipate a sufficient amount of energy to become a bound binary in the tidal field of the central SMBH. Jacobi capture is an efficient channel for binary BHs formation in AGN discs \citep{2022arXiv220309646B}. 

However, the formation of binary WDs via gravitational wave emission is difficult. The physical radius of WDs is around $\sim 10^4 {\rm km}$, which is three orders of magnitude larger than the Schwarzschild radius of stellar masssive BHs. During the  close encounter of two WDs, the separation for strong gravitational emission  is much smaller than the physical radius of WDs. What happens in a close encounter is therefore the collision of two WDs rather than the formation of a binary WDs. These white dwarf collisions, which could produce the type Ia supernovae \citep{2009ApJ...705L.128R, 2009MNRAS.399L.156R, 2012ApJ...759...39H}, have been studied in field hierarchical systems \citep{2018MNRAS.478..620H} and globular clusters \citep{2010ApJ...724..111R}. Compared with these environment, the high-density of WDs in AGN discs makes the WD-WD collisions happen.

In this work, we focus on the WD-WD collisions in the restricted three-body systems. Such systems consist with two WDs orbiting around a central SMBH; the separation between two WDs orbits is of the order of mutual Hill radii. Since the system is dynamically unstable, the two WDs may collide during the close encounters. We perform a series of $N$-body simulations to study the closest separation and WD-WD collisions in close encounters. Each set of simulations has different initial parameters: the initial orbital separation, the relative orbital inclination, and the mass of the central SMBH. The effect of different initial orbital radius is also considered.

The rest of the paper is organized as follows. In Section~\ref{Section 2}, we describe our analytical framework and initial numerical setups. In Section~\ref{section3}, we present the distribution and time evolution of closest separation between WDs and the fraction of WD collision. Finally, the main conclusions and discussion are summarized in Section~\ref{section4}.

\section{Methods}
\label{Section 2}
A large population of WDs is expected to be embedded in AGN accretion discs as a result of 
stellar evolution and dynamical friction. Such WDs finally circulate around the central SMBH with close orbits, due to the alignment and migration of WDs. Whilst WDs have close circular orbits, they may experience very close encounters, during which collision can happen. These direct collision between two WDs may lead to type Ia supernovae \citep{2018MNRAS.478..620H, 2010ApJ...724..111R, 2021MNRAS.507..156G}. 

As the WDs align with the discs, gas torques might cause the WD orbits to change over time as they undergo the so called type I migration within the discs. 
The type I migration timescale \citep{Tanaka_2002, 2010MNRAS.401.1950P, 2012MNRAS.425..460M, 2018ApJ...866...66M} can be estimated as
\begin{equation}
\begin{aligned}
t_{\rm{mig}} \approx &{\rm{7.9Myr}} \left(\frac{N}{3}\right)^{-1} \left(\frac{r}{100r_{\rm g}}\right)^{-\frac{1}{2}} \left(\frac{m}{0.6M_{\odot}}\right)^{-1} \\
                         &\times \left(\frac{h/r}{0.01}\right)^2 \left(\frac{\Sigma}{10^5 \rm{kg\ m^{-2}}}\right)^{-1} \left(\frac{M}{10^6M_{\odot}}\right)^{\frac{3}{2}},
\end{aligned}
\label{eq:1}
\end{equation}
where $m$ and $M$ are the mass of the WD and the central SMBH respectively, $r$ is the semi-major axis of WD, $r_{\rm g}$ is the gravitational radius, defined as $GM/c^2$, 
$\Sigma$ is the surface density of the disc, $h/r$ is the disc aspect ratio, and $N$ is a numerical factor of the order of 3. The lifetime of AGN disc is around $10 {\rm Myr}$. For a dense Sirko \& Goodman \citep{2003MNRAS.341..501S} model disc with $M = 10^8M_{\odot}$, a WD migrate to $100r_{\rm g}$ in $\sim 2.5 {\rm Myr}$ \citep{2020MNRAS.494.1203M}. Consequently, it is possible for a WD to migrate to the inner AGN discs. In addition to the type I migration, WDs can also exist in inner AGN discs if those WDs  align with the discs at small orbits. As the WDs migrate inward, they finally collect in the migration trap, where the net torque on a migrator is zero \citep{2016ApJ...819L..17B}. Due to their different radii of alignment, migration rate, and their collection in the migration trap, WDs may have very close orbits in the AGN discs.

\begin{figure}
\centering
\includegraphics[width=0.8\columnwidth]{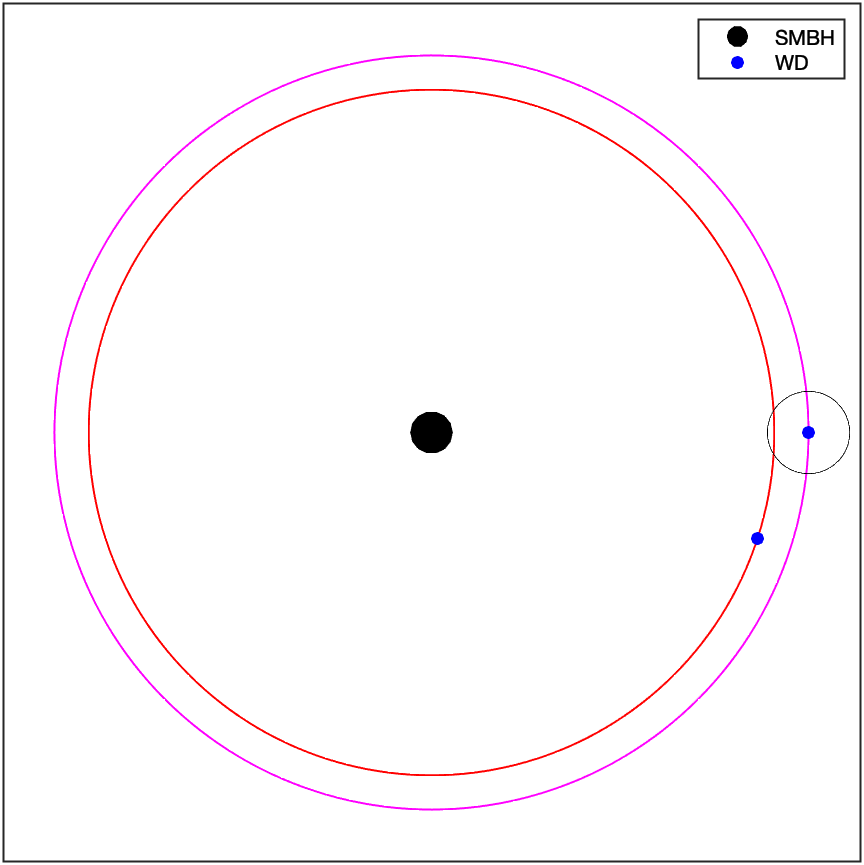}
\vspace*{-1mm}
\caption{Picture of the simulation system. The system consist with a central SMBH (black dot) and two WDs (blue dot). The two WDs orbit on nearly circular and nearly co-planer orbits around the SMBH.}
\label{figure1}
\end{figure}

We consider a system consisting with a  central SMBH, orbited by two WDs in nearly circular and nearly co-planer orbits (see Figure~\ref{figure1}). Due to their alignment \citep{2019ApJ...876..122Y, 2020MNRAS.499.2608F} and migration \citep{Tanaka_2002, 2010MNRAS.401.1950P}, the orbits of the two WDs may be very close to each other. In this paper, we set the central SMBH mass as $M$, the two WDs masses as $m_1$, $m_2$, and the initial semi-major axis around the SMBH as $a_1$, $a_2$. We define $p = a_2 - a_1$ to be the initial orbital separation. Different initial semi-major axis cause the two WDs to have slightly different orbital periods. In consequence, their relative separation gradually decreases with time. If their initial orbital separation $p$ is much larger than their mutual Hill radius, their orbits are still stable. If their initial orbital separation $p$ is of the order of their mutual Hill radius, their orbits might become unstable and the subsequent evolution become chaotic. Here, the mutual Hill radius is defined as 

\begin{equation}
R_{\rm H} \equiv \frac{a_1 + a_2}{2} \left( \frac{m_1 + m_2}{3M} \right)^{1/3} \approx a_1 \left( \frac{m_1 + m_2}{3M} \right)^{1/3}.
\label{eq:2}
\end{equation} 
If we ignore the influence of the disc, the boundary between "stable" and "unstable" gives  as a critical orbital
separation $p_{\rm c}$ \citep{1993Icar..106..247G}
\begin{equation}
p_{\rm c} = 2\cdot 3^{1/6}\left( \frac{m_1 +m_2}{3M} \right)^{1/3} = 2\sqrt{3} R_{\rm H}.
\label{eq:3}
\end{equation}
While $p < p_{\rm c}$, the orbital evolution will become chaotic and Jacobi captures start to play an important role during the 
close encounters.

As a result, the closest separations of the two WDs can reach small values during close encounters. 
The bonding energy of a stable binary WD at the Hill radius is given by

\begin{equation}
E_{\rm b} = \frac{Gm_1 m_2}{2R_{\rm H}}.
\label{eq:4}
\end{equation}
At the closest separation, the energy dissipation by gravitational wave emission can be given 
by \citep{PhysRev.136.B1224, 1977ApJ...216..610T}
\begin{equation}
  \Delta E_{\rm{GW}} = \frac{85\pi}{12\sqrt{2}} \frac{G^{7/2} (m_1 m_2)^2 (m_1 + m_2)^{1/2}}{c^5 r_{\rm p}^{7/2}},
  \label{eq:5}
\end{equation}
where $r_{\rm p}$ is the closest separation of the two WDs. If we neglect the possible effect of tidal dissipation and gas drag, the gravitational wave emission is the only mechanism 
to dissipate energy between two WDs. To form a stable binary WD, we need $\Delta E_{\rm GW} \gtrsim E_{\rm b}$ \citep{2022arXiv220305584L}, i.e. 

\begin{equation}
r_{\rm p} \lesssim r_{\rm b} \equiv 3.48\left(\frac{m_1 m_2}{(m_1 + m_2)^2} \right)^{\frac{2}{7}} \left(\frac{m_1 + m_2}{M} \right)^{\frac{10}{21}} \left( \frac{GM/c^2}{a_1} \right)^{\frac{5}{7}} R_{\rm H}.
\label{eq:6}
\end{equation} 
Therefore, we can probe for a critical separation value of $r_{\rm b}$, below which a bonding binary WD can be formed by 
gravitational wave emission with given $M$ and $a_1$. To eject one of the two WDs from the system during close encounters, we require 
\begin{equation}
\frac{Gm_1 m_2}{r_{\rm e}} \gtrsim \frac{1}{2} \frac{m_1 m_2}{m_1 + m_2}v_{\rm orb}^2
\label{eq:7}
\end{equation}
where $r_{\rm e}$ is the ejection separation and $v_{\rm orb}$ is the orbital velocity, $v_{\rm orb} = \sqrt{GM/a_1}$. For two WDs with mass $m_1 = m_2 = 0.6M_{\odot}$, we define the 
direct WD-WD collision separation as the sum of the physical radii of the WDs, $r_{\rm c} = 2r_{_{\rm WD}}$, which is approximate $1.5\times 10^7{\rm m}$. 
If $M = 10^6M_{\odot}$, $a_1 = 100GM/c^2$, we find a critical gravitational wave bonding radius $r_{\rm b} \lesssim 1.76\times 10^5{\rm m}$, 
which is much smaller than the direct collision separation $r_{\rm c} \approx 1.5\times 10^7 {\rm m}$. The ejection separation $r_{\rm e} \lesssim 3.6\times 10^5{\rm m}$.
The two WDs therefore collide together before they form a binary WD through the emission of gravitational wave or they might be ejected from the system. 
As a result, there are two possible outcomes for WD close encounters: (i) the two WDs experience a relatively soft close encounter and the orbits separate to "stable" orbits; (ii) The two WDs experience 
a very close encounter and collide with each other.

\begin{table}
  \centering
  \caption{The initial input parameters. The square bracket represent the uniform distribution between the two value in 
    square bracket. Column 1 is the run name. Column 2 is the mass of SMBH in N-body units. Column 3 is the mass of WDs in N-body units. 
    The two WDs have equal mass .Column 4 is the initial orbit radius in N-body units. Column 5 shows the initial orbital 
    separations and Column 6 the number of samples.}
  \label{table1}
  \begin{tabular}{lcccccc}
    \hline
    name & $M$ & $m_1,m_2$ & $a_1$ & $p/R_{\rm H}$ & $i_1,i_2$ & $N$\\
    \hline
    \hline
    Run1 & 1 & $6\times 10^{-7}$  & 1 & 1.0 & 0 & 1000\\
    \hline
    Run2 & 1 & $6\times 10^{-7}$ & 1 & 1.5 & 0 & 1000\\
    \hline
    Run3 & 1 & $6\times 10^{-7}$ & 1 & 2.0 & 0 & 1000\\
    \hline
    Run4 & 1 & $6\times 10^{-7}$ & 1 & 2.5 & 0 & 1000\\
    \hline
    Run5 & 1 & $6\times 10^{-7}$ & 1 & 3.0 & 0 & 1000\\
    \hline
    Run6 & 1 & $6\times 10^{-7}$ & 1 & 3.5 & 0 & 1000\\
    \hline
    Run7 & 1 & $6\times 10^{-7}$ & 1 &  [0.8,4] & 0 & 4000\\
    \hline
    Run8 & 1 & $6\times 10^{-7}$ & 1 &  3.0 & 
    $|i_1 - i_2{|}{=}10^{-3}\frac{R_{\rm H}}{a_1}$ & 1000\\
    \hline
    Run9 & 1 & $6\times 10^{-7}$ & 1 &  3.0 & 
    $|i_1 - i_2{|}{=}10^{-2}\frac{R_{\rm H}}{a_1}$ & 1000\\
    \hline
    Run10 & 1 & $6\times 10^{-7}$ & 1 &  3.0 & 
    $|i_1 - i_2{|}{=}10^{-1}\frac{R_{\rm H}}{a_1}$ & 1000\\
    \hline
    Run11& 1 & $6\times 10^{-7}$ & 1 &  3.0 & 
    $|i_1 - i_2{|}{=}\frac{R_{\rm H}}{a_1}$ & 1000\\
    \hline
    Run12 & 1 & $6\times 10^{-8}$ & 1 & 3.0 & 0 & 1000\\
     \hline
    Run13 & 1 & $6\times 10^{-9}$ & 1 & 3.0 & 0 & 1000\\
    \hline
  \end{tabular}
  %\vspace*{-8mm}
\end{table}   

In our simulations, we adopt N-body units in which $G = 1$, central SMBH mass $M = 1$, the WD orbit $a_1 = 1$. So we can 
get the initial period $P_1 = 2\pi$. The WD masses in units of the mass of SMBH are $m_1 = m_2 = 6\times 10^{-7}$ and 
 $m_1 = m_2 = 6\times 10^{-8}$ for SMBH masses of  $10^6M_{\odot}$ and $10^7M_{\odot}$ respectively (i.e. $m_1 = m_2 = 0.6M_{\odot}$). 
We note that $r_{\rm g} \approx 1.48 \times 10^9{\rm m}$ and $1.48\times 10^{10}{\rm m}$ for $M = 10^6M_{\odot}$ and $10^7M_{\odot}$ 
respectively, and the WD-WD collision separation is $r_{\rm c} \approx 1.5 \times 10^7{\rm m}$. For convenience, we take the corresponding 
WD-WD collision separation to be $10^{-2}r_{\rm g}$ and $10^{-3}r_{\rm g}$ respectively. The initial orbital separation $p$ is set to a series 
of values from $0.8R_{\rm H}$ to $4.0R_{\rm H}$. The two WDs orbits have initial eccentricities $e_1 = 0$, $e_2 = 10^{-5}$. The initial 
difference of the longitude is uniform distributed in the range $[0,2\pi]$. The two WDs orbits may not be in the same plane. We define the inclination angle 
between the orbital plane and the disc plane as $i_1$ and $i_2$. All details of the initial parameters are summarized in Table~\ref{table1}.

In this study, we integrate the three-body systems using the N-body code REBOUND \citep{2012A&A...537A.128R} with the IAS15 
integrator \citep{1985ASSL..115..185E, 2015MNRAS.446.1424R}. For each individual simulation, we run $10^5P_1$. 

\section{Results}
\label{section3}
In this section we present our simulation results with different initial parameters. In the following analysis, we consider the different initial orbital radius $a_1$ 
with physics values rather than N-body units. We record the first time at which the separation between two WDs is smaller than $r_{\rm c}$ as the collision time. In Section~\ref{section3.1} we present the closest separation between two WDs and WD-WD collisions as a function of the initial orbital separation. In Section~\ref{section3.2}, we show the influence of the relative orbital inclination on system close encounters and WD-WD collisions, 
and consider the mass of the central SMBH in section~\ref{section3.3}. 
Finally, the event rate of the WD-WD collision in AGN discs is discussed in 
Section~\ref{section3.4}.

\subsection{Closest separation and WD-WD Collisions}
\label{section3.1}
The close encounter criterion in this work is that the separation between WDs should be much less than the initial orbital separations, i.e. 
$\Delta{r} = |\bm{r_1}-\bm{r_2}| \lesssim 10^{-1}p$. Here we consider the close encounters between two WDs in different initial orbital separation: the closest separation and collisions. WD-WD collision occurs when the closest separation is smaller than the WD-WD 
collision separation. We also discuss the WD-WD collision fraction at different initial orbit $a_1$.

\begin{figure*}
\centering
\includegraphics[width=2\columnwidth]{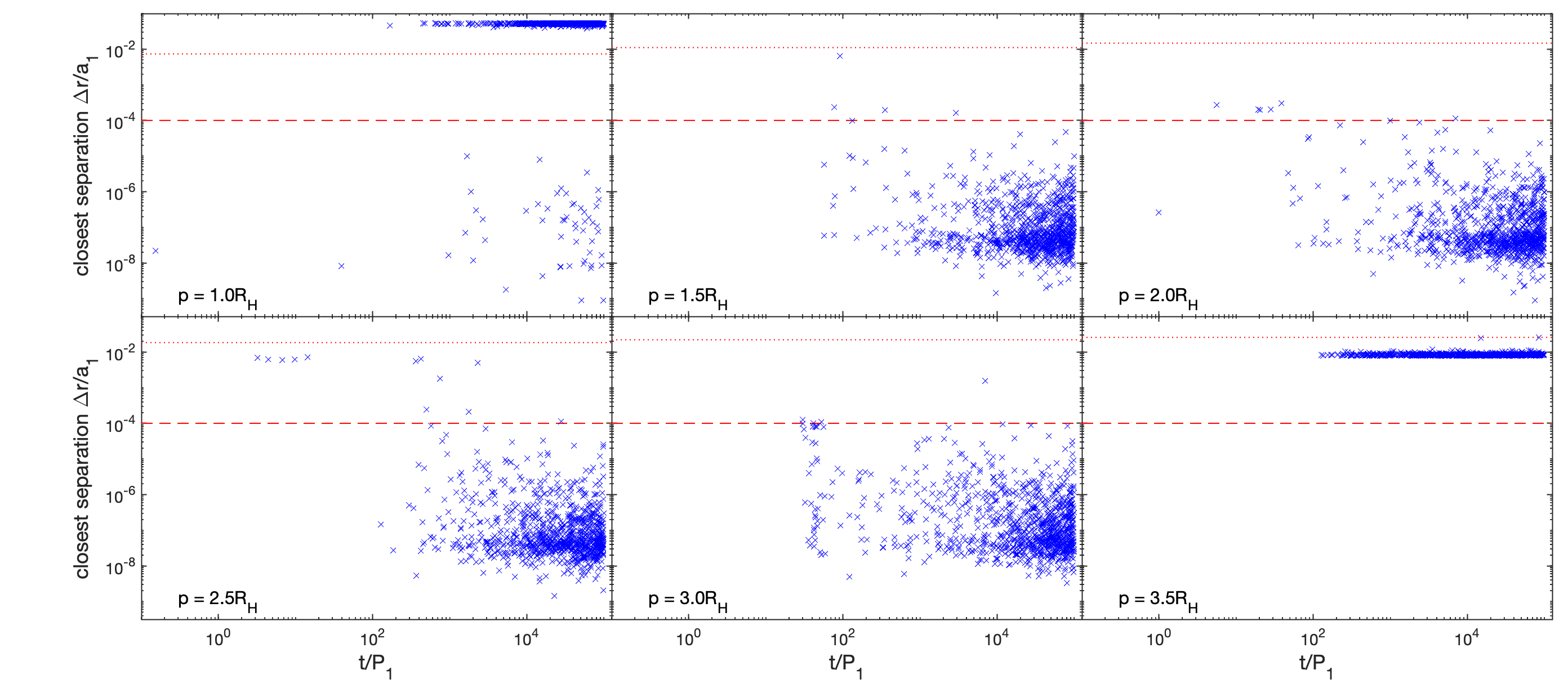}
\vspace*{-3mm}
\caption{The closest separation between two WDs as a function of time at the closest separation. The blue crosses are individual simulation run. 
The initial orbital separation is $p = 1.0$, $1.5$, $2.0$, $2.5$, $3.0$, $3.5R_{\rm H}$. There are 1000 runs for every $p$.  The red dashed 
line indicate the WD-WD collision separation with $a_1 = 100r_{\rm g}$. The read dotted lines represent $p = \Delta{r}$.}
\label{figure2}
\end{figure*}

In Fig.~\ref{figure2}, we plot the resulting closest separation in Run1-Run6. The blue crosses correspond to simulation samples with the  
given initial parameters. The red dotted lines show the initial orbital separation. Equation~(\ref{eq:3}) gives the boundary between 
dynamically "stable" and "unstable" regions, criterion $p_{\rm c} = 2\sqrt{3}R_{\rm H}$. For $p = 3.5R_{\rm H} > 2\sqrt{3} R_{\rm H}$ in Run6, 
the closest separation is slightly smaller than the initial orbital separation and there are no close encounters. Conversely, Run2-Run5
with the unstable orbits undergo close encounters, and the closest separation is much smaller than the initial orbital separation. Run1 also have "unstable" initial orbital separation, but only around 5 per cent of samples undergo close encounters. The red dashed line corresponds to a WD-WD collision separation corresponding to $a_1 = 100r_{\rm g}$. We see that WD-WD collisions occurred in nearly all samples for $p = 1.5,2.0,2.5,3.0R_{\rm H}$, but only a small part for $p = 1.0R_{\rm H}$, and no collision occurred for $p = 3.5R_{\rm H}$.

\begin{figure}
\centering
\includegraphics[width=\columnwidth]{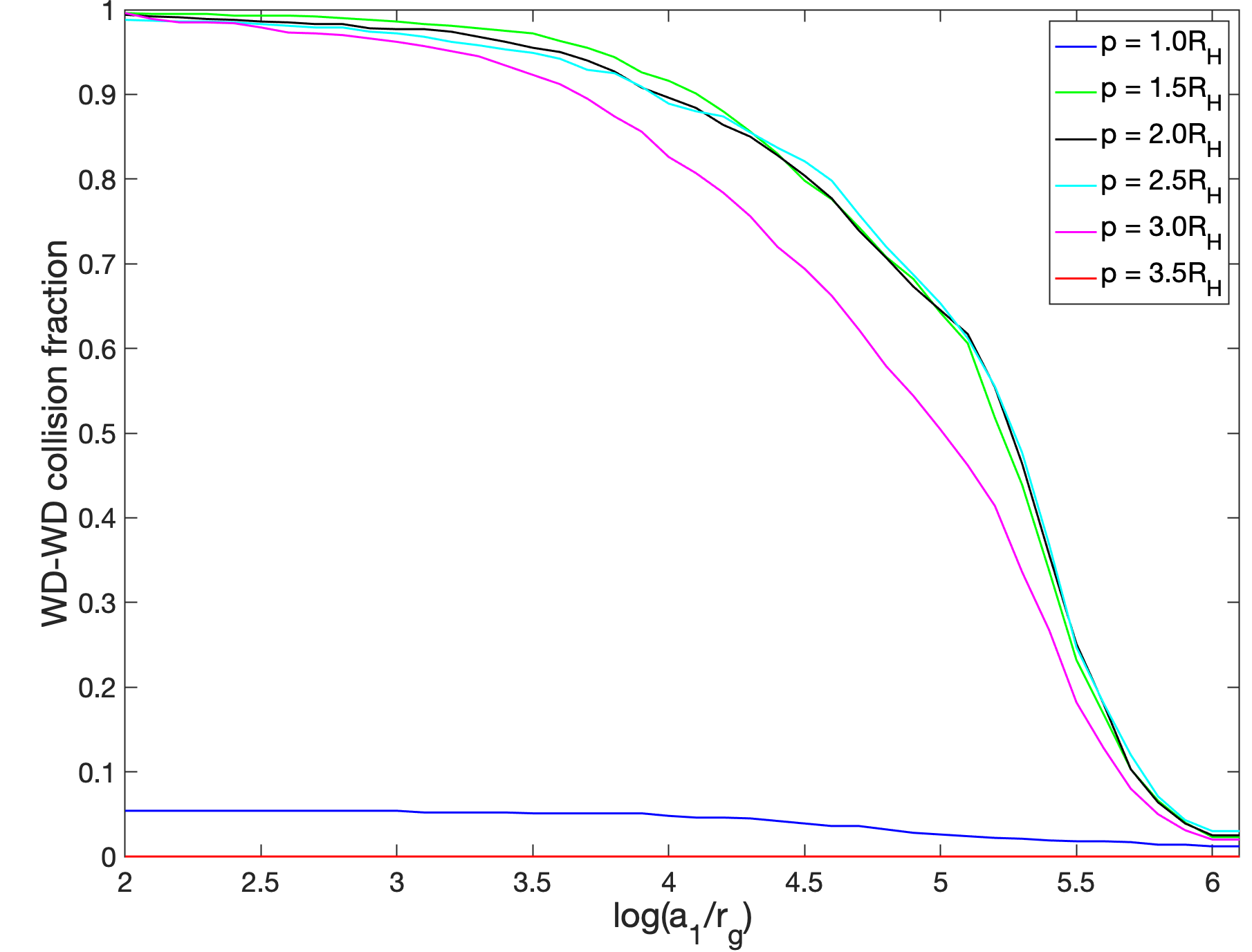}
\vspace*{-3mm}
\caption{The WD-WD collision fraction as a function of initial WD orbital semi-major radius $a_1$. Different colors correspond to the different initial orbital separation plot 
in Figure~\ref{figure1}. We can see that the WD-WD collision fraction decrease as the $a_1$ increase. For $p = 3.5R_{\rm H}$ and $p = 1.0R_{\rm H}$, the collision fraction is much smaller than others, and there is no WD-WD collision for $p = 3.5R_{\rm H}$.}
\label{figure3}
\end{figure}

We can define the WD-WD collision separation in units of $a_1$. The relative collision value $r_{\rm c}/a_1$ decreases as $a_1$ increases. As a result,  
the collision fraction decreases. Fig.~\ref{figure3} shows the collision fraction in cases with  different initial orbit $a_1$. For $p = 3.5R_{\rm H}$, which means stable 
orbit evolution, there is no collision occurring. While $p < p_{\rm c}$, the WD orbits are unstable and the collision fraction decrease as $a_1$ increase. 
As  the initial orbit $a_1$ changes, the WD-WD collision fraction decreases slowly at $a_1 < 10^4r_{\rm g}$, 
but when $a_1 > 10^4r_{\rm g}$ the WD-WD collision fraction decreases quickly to 50 per cent at $10^5r_{\rm g}$ and 2 per cent at $10^6r_{\rm g}$. From Equation~(\ref{eq:7}), it can be seen that, at the orbital radii $a_1 \gtrsim 10^4r_{\rm g}$, the ejection radius $r_{\rm e} \simeq 3.6\times 10^7{\rm m} > r_{\rm c}$, which means one of the WDs may be ejected before they collide.  In Run1, only 5 per cent of all samples show the WD-WD collision for $p = 1.0R_{\rm H}$, the reason will be given next.

\begin{figure}
\centering
\includegraphics[width=\columnwidth]{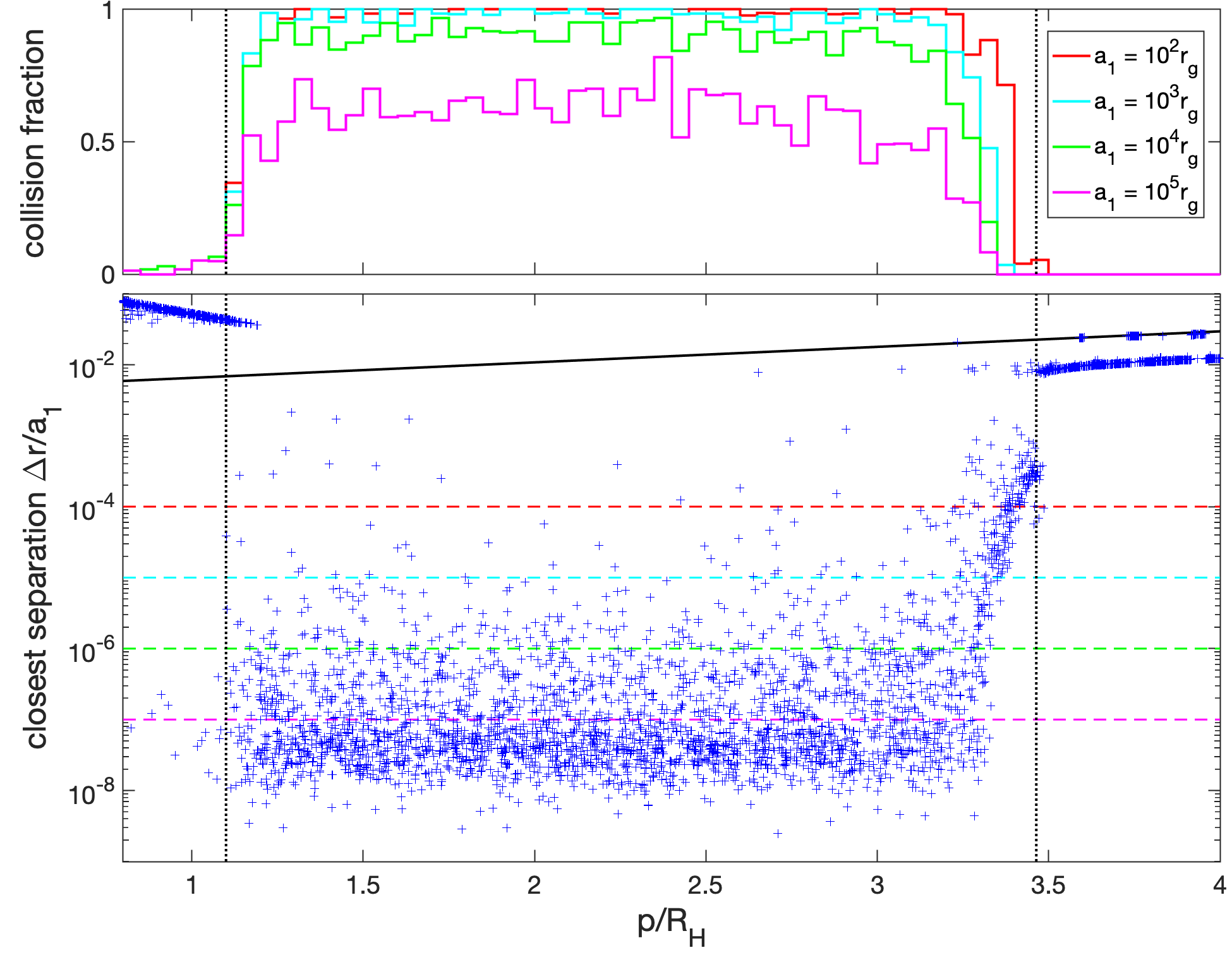}
\vspace*{-3mm}
\caption{\textbf{Upper panel:} WD-WD collision fraction as a function of initial orbital separation at a given initial orbit $a_1$. Different color correspond 
    to different initial $a_1$. \textbf{Bottom panel:} closest separation as a function of initial orbital separation. The brown dotted line corresponds to $\Delta{r} = p$. 
    The red, cyan, green and magenta dashed line are the WD-WD collision separation of $a_1 = 10^2,10^3,10^4,10^5r_{\rm g}$ respectively. The black 
    dotted line is the boundary between "stable" and "unstable" $p_{\rm c} = 2\sqrt{3}R_{\rm H}$.}
\label{figure4}
\end{figure}

\begin{figure}
\centering
\includegraphics[width=\columnwidth]{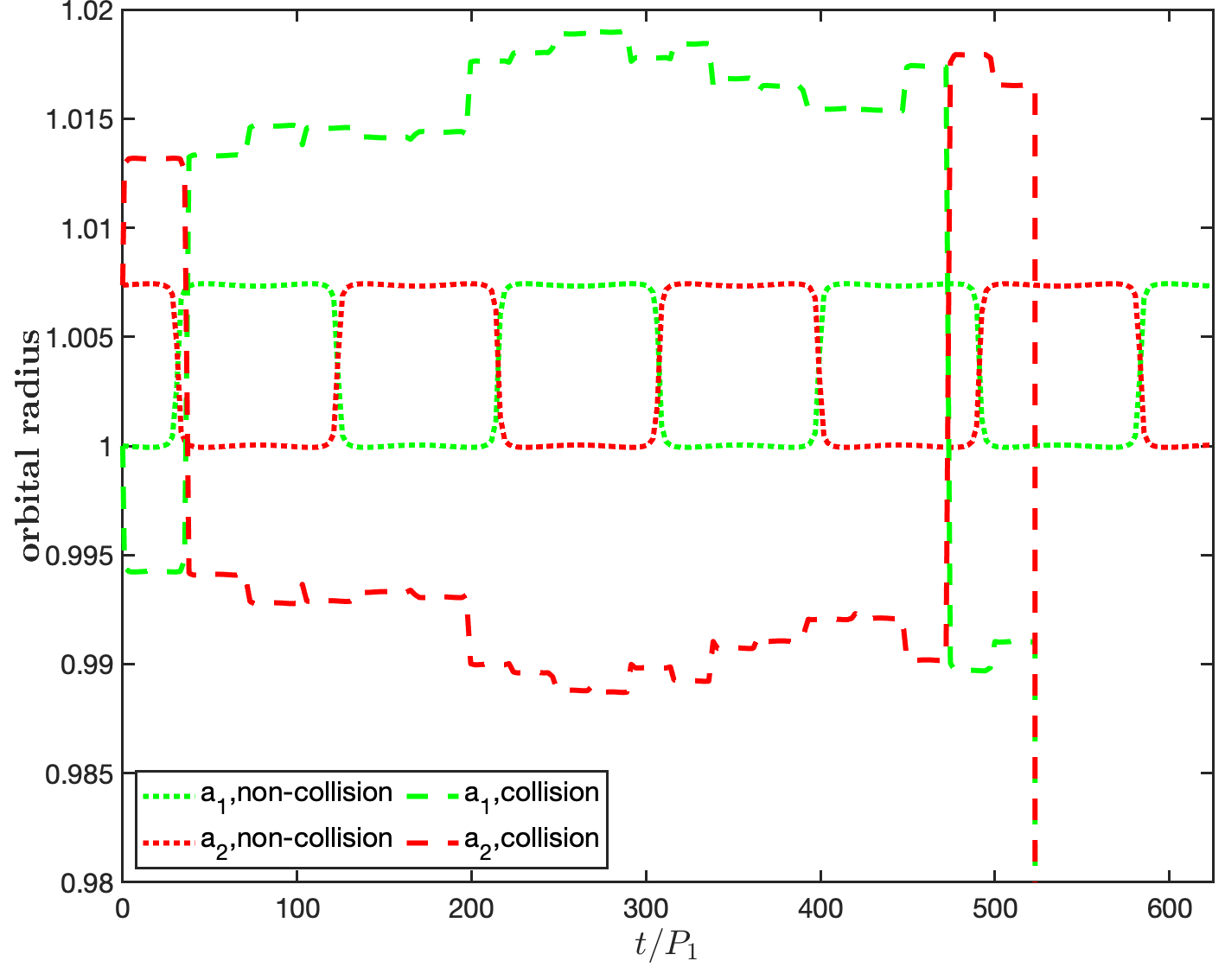}
\vspace*{-3mm}
\caption{Orbits evolution of WDs at $p=R_{\rm H}$. The dotted curves represent non-close encounter orbits and the dashed curves represent close encounter orbits. The color red and green correspond to $a_1$ and $a_2$ respectively.}
\label{figure5}
\end{figure}

Fig.~\ref{figure4} shows the closest separation and WD-WD collision fraction in Run7 ($p/R_{\rm H} \in [0.8,4]$). The bottom panel plots the closest separation as a function of 
initial orbital separation. The black dotted line shows the result for $p = p_{\rm c}$. The red, cyan, green and magenta dashed lines show the WD-WD collision separation for the cases $a_1 = 10^2,10^3,10^4,10^5r_{\rm g}$ respectively. We find that almost all close encounters take place in the  dynamical "unstable" region (i.e. $p < p_{\rm c}$).  However, WD-WD do not undergo a close encounter for $p < p_{\rm c}$ in all samples. We can see that, at $p \lesssim 1.1R_{\rm H}$, WD-WD close encounters become rarely, and most of  the closest separation increases as the initial orbital separation decreases. The reason can be explained as follows. At small initial separation, the WDs are located on a horseshoe or a tadpole orbit in the rotating frame of reference. As they approach each other, one WD orbit decreases and another one increases, which leads to the exchange of their orbits. However, a few samples have also experienced close encounters, because their initial separation is very close, which lead to the initial orbit being outside the horseshoe orbit. Fig.~\ref{figure5} shows an example of the evolution of the orbital radius in the close encounter case and non-close encounter case for $p=R_{\rm H}$. We can see that the two WDs exchange their orbits every time they encounter each other in the non-close encounter case, but in the close encounter case, the evolution of the orbits are chaotic and one WD is ejected in a very close encounter at around $520P_1$. 

\begin{figure}
\centering
\includegraphics[width=\columnwidth]{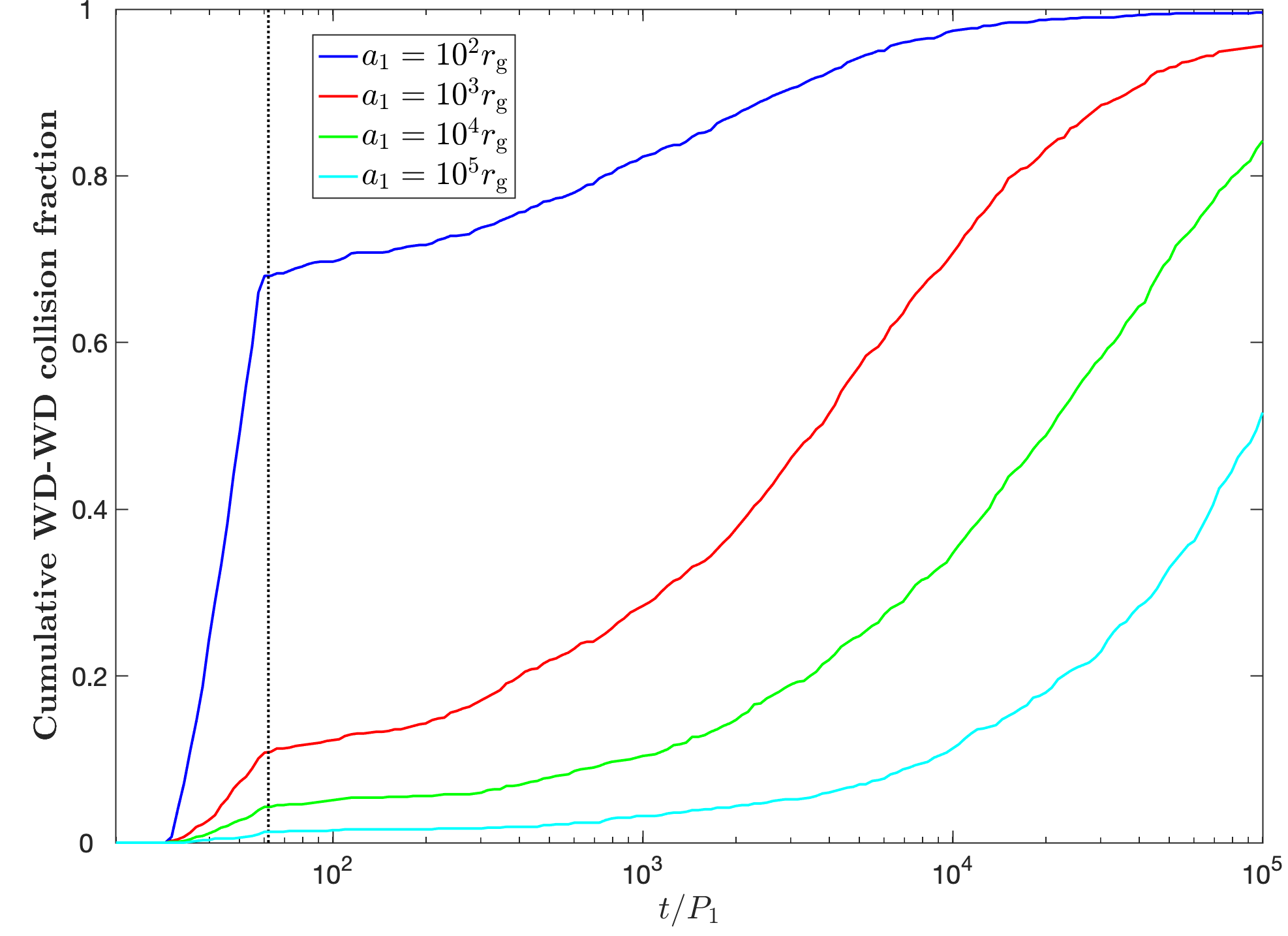}
\vspace*{-3mm}
\caption{The cumulative WD-WD collision fraction distribute as a function of simulation time for Run5. The colored curves correspond to the initial orbit radius $a_1=10^2,10^3,10^4,10^5r_{\rm g}$ respectively. The black dotted line represents time $t = 62P_1$. }
\label{figure6}
\end{figure}

Fig.~\ref{figure6} shows the cumulative WD-WD collision fraction as a function of time in Run5 (i.e. $p = 3.0R_{\rm H}$). The colored solid lines correspond to the results for the different initial orbital radii $a_1$. The black dotted line represents the time $t = 62P_1$, which is twice the lowest common multiple of $P_1$ and $P_2$. We find that, at the initial orbital radius $a_1=10^2r_{\rm g}$, WD-WD collisions occur in almost all samples. As the initial orbital radius increases, the WD-WD collision fraction decrease. For instance, in the case of $a_1 = 10^5r_{\rm g}$,
only half the samples in our runs show WD-WD collisions.
For the WD-WD collision samples, around 70 per cent of WD-WD collisions occur during the first two encounters (i.e. $t < 62P_1$) for $a_1 = 10^2r_{\rm g}$, but only 30 per cent at $t > 62P_1$. As the orbital radii increase, WD-WD collisions occurring during the first two encounters decrease sharply. 

Due to the type I migration of WDs, the orbital radii will decrease. 
The time-scale of type I migration can be estimated from equation~(\ref{eq:1}). Here we simply estimate the WD-WD collision timescale as $t_{\rm c} = 10^5P_1/f_{\rm c}$, where $f_{\rm c}$ is the cumulative WD-WD fraction in $10^5P_1$. In the outer region ($r \gtrsim 10^5r_{\rm g}$) of the discs, the type I migration time-scale is smaller than the WD-WD collision timescale, i.e. $t_{\rm mig}(r) \lesssim t_{\rm c}(r)$, which means the two WDs may migrate to the inner region ($r < 10^5r_{\rm g}$) before they collide. The large ejection separation in the outer region may lead to one of the WDs being ejected from the system, which can also decrease the rate of WD-WD collision. In conclusion, WD-WD collisions mainly occur in the inner region ($r < 10^5r_{\rm g}$) of the discs; they are rare in the outer region because most of the WDs have migrated into the inner region or been ejected when they collide.

\subsection{Results with initial inclinations}
\label{section3.2}
The existence of an initial relative orbital inclinations can affect the orbital separation between two WDs. Assuming that two WDs have a small relative inclination $\Delta{i} = |i_1-i_2|$, the orbital distance at the same longitude is larger than that without inclination. Naturally, we expect a decrease in close encounter fraction as the relative inclination $\Delta{i}$ increases. Consequently, the closest separation becomes larger and the collision fraction will decrease. Run8 - Run12 are a series of simulations with the different inclinations.

\begin{figure}
\centering
\includegraphics[width=\columnwidth]{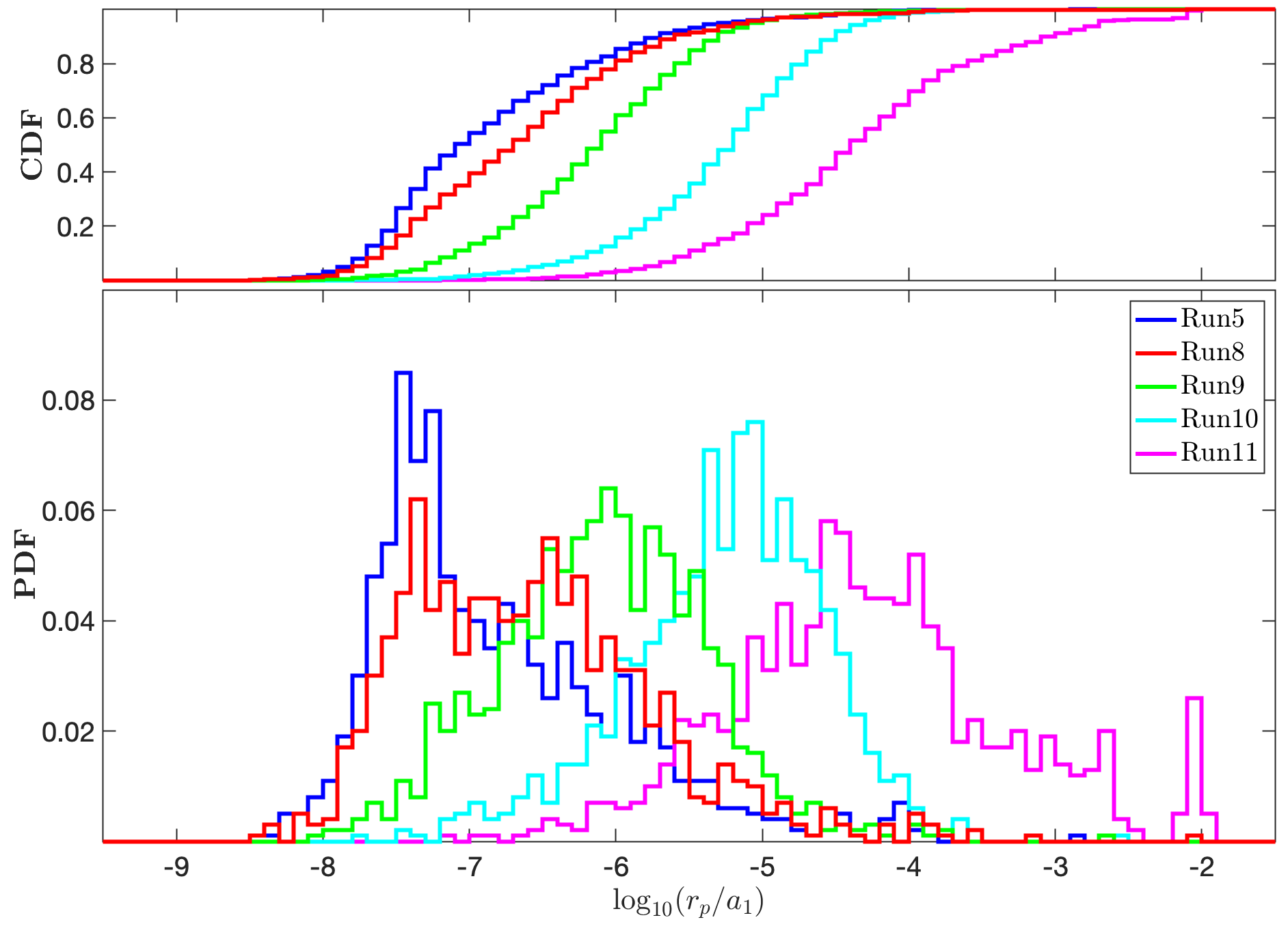}
\vspace*{-4mm}
\caption{Probability density function of the closest separation between two WDs in close encounters. The colored curves correspond coplanar orbits (blue curve), $\Delta{i} = 10^{-3}R_{\rm H}/a_1$ (red curve), $\Delta{i} = 10^{-2}R_{\rm H}/a_1$ (green curve), $\Delta{i} = 10^{-1}R_{\rm H}/a_1$ (cyan curve), and $\Delta{i} = R_{\rm H}/a_1$ (magenta curve).}
\label{figure7}
\end{figure}

Fig.~\ref{figure7} shows the distribution of probability of closest separation $r_{\rm p}$ between the two WDs. The upper panel shows the cumulative distribution and the bottom panel the probability density function. The colored lines correspond to   results with different initial inclinations. We find that the peak of the closest separation for coplanar orbits is around $10^{-7.5}a_1$, and nearly all closest separation have $r_{\rm p} \lesssim 10^{-5}a_1$. As the relative inclination $\Delta{i}$ increases, the peak of the closest separation increases. For small relative inclination $\Delta{i} \lesssim 10^{-1}R_{\rm H}/a_1$, almost all of the closest separation have $r_{\rm p} \lesssim 10^{-4}a_1$. However, for $\Delta{i} = R_{\rm H}/a_1$, around 20 per cent of the samples with closest separation $r_{\rm p} > 10^{-4}a_1$. 

\begin{figure}
\centering
\includegraphics[width=\columnwidth]{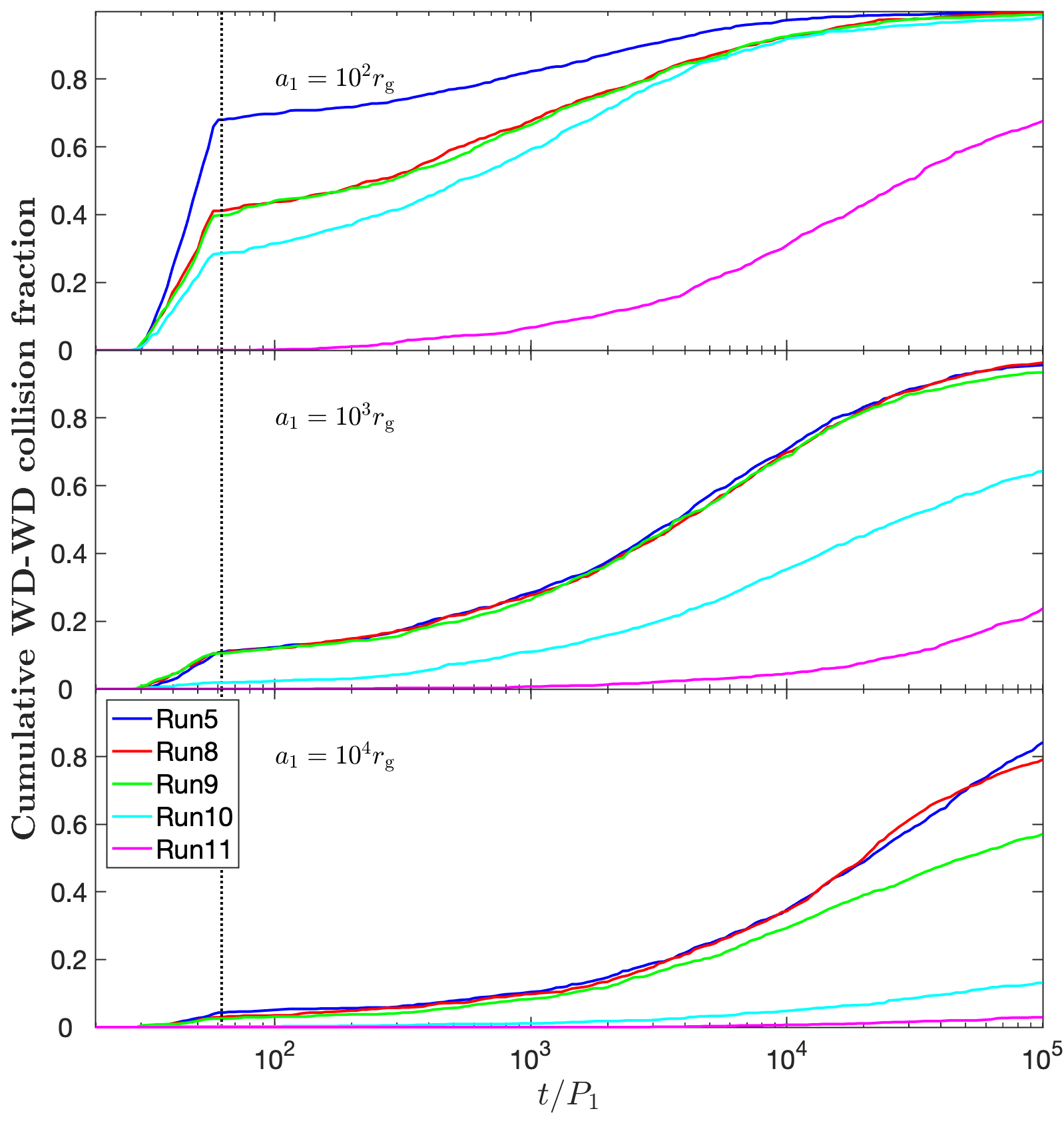}
\vspace*{-4mm}
\caption{The cumulative WD-WD collision fraction as a function of time. From the top panel to bottom correspond to the initial orbital radius $a_1 = 10^2r_{\rm g}$, $10^3r_{\rm g}$, $10^4r_{\rm g}$. The colored curves correspond to different initial inclination given in Table~\ref{table1}. The black dotted line represents time $t = 62P_1$.}
\label{figure8}
\end{figure}

Fig.~\ref{figure8} shows the cumulative WD-WD collision fraction as a function of time for the different relative inclinations. From the top panel to the bottom one, the initial orbital radius $a_1$  is set to be $10^2r_{\rm g}$,  $10^3r_{\rm g}$ and $10^4r_{\rm g}$, respectively. We find that, for the relative inclinations $\Delta{i} \lesssim 10^{-2}R_{\rm H}/a_1$, the cumulative WD-WD collision fraction is almost the same as that in the case of coplanar orbits at $10^5P_1$, but for large relative inclinations $\Delta{i} > 10^{-2}R_{\rm H}/a_1$, the WD-WD collision fraction decreases quickly, and the larger the relative inclination, the faster the WD-WD collision fraction decreases. This indicates that small relative inclinations have nearly no effect on WD-WD collision, but large relative inclination can reduce the WD-WD collision rate sharply. There are almost no WD-WD collision for $\Delta{i}=R_{\rm H}/a_1$ at all radius in the first $62P_1$. However, for $\Delta{i} \lesssim 10^{-2}R_{\rm H}/a_1$, the WD-WD collisions are considerable ($\gtrsim$ 10 per cent) in the first $62P_1$ for $a_1 \lesssim 10^3r_{\rm g}$. For $\Delta{i}=10^{-1}R_{\rm H}/a_1$, around 30 per cent of WD-WD collisions occur in first $62P_1$ for $a_1 = 10^2r_{\rm g}$, but there are almost no WD-WD collisions for $a_1\gtrsim 10^3r_{\rm g}$.

\subsection{Results with the different mass of SMBH}
\label{section3.3}
The mass of the central SMBH can span several orders of magnitude. In Run13 and Run14, we investigate the effects of SMBH mass $M=10^7M_{\odot}$ and $10^8M_{\odot}$, corresponding to mass ratio between WDs and SMBH of $6\times 10^{-8}$ and $6\times 10^{-9}$ respectively.

\begin{figure}
\centering
\includegraphics[width=\columnwidth]{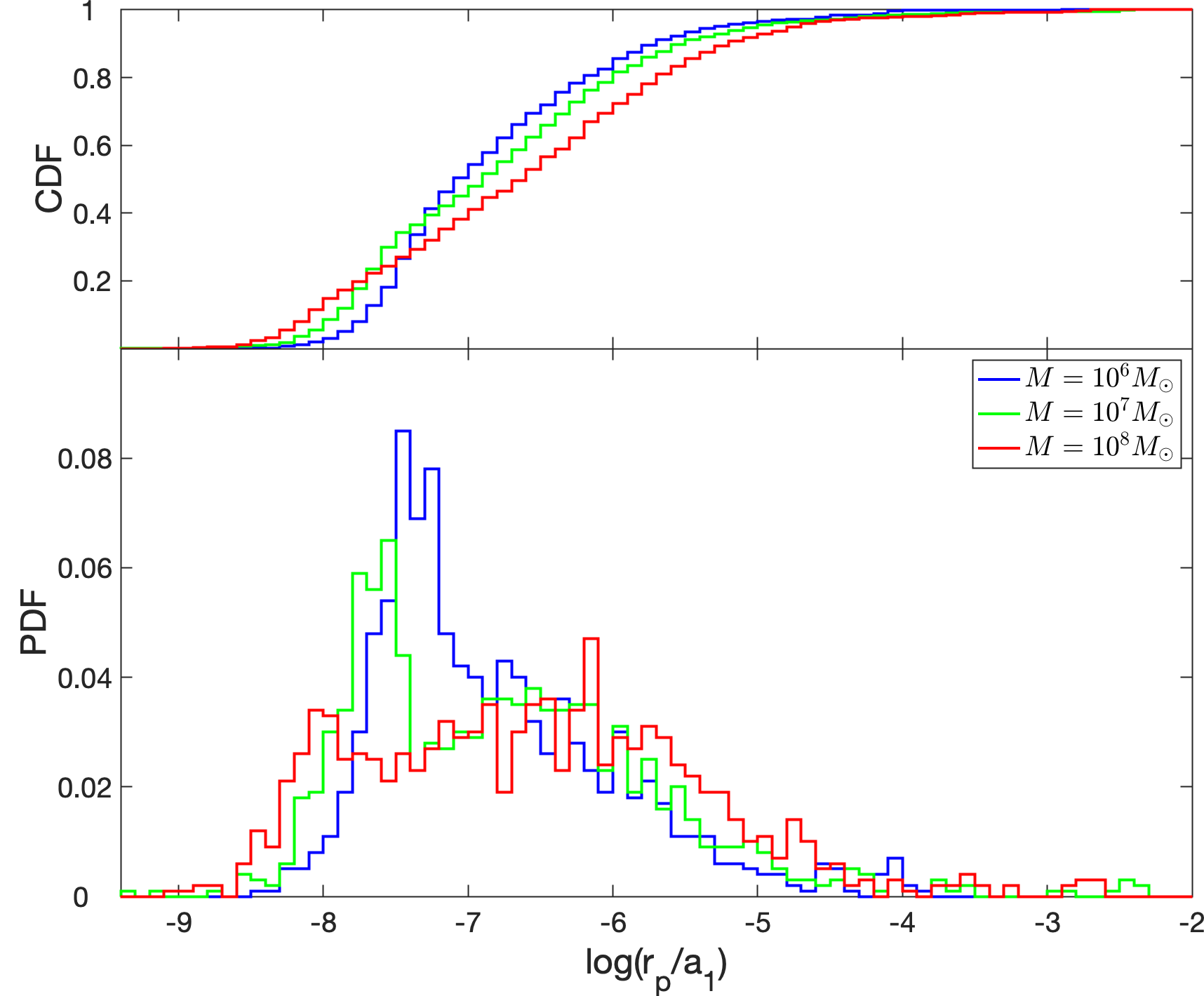}
\vspace*{-4mm}
\caption{\textbf{Upper panel:} Cumulative closest separation distribution as a function of closest separation. \textbf{Bottom panel:} Probability distribution of closest separation as a function of closest separation. The blue, green and red curve correspond to the central SMBH mass $M = 10^6M_{\odot}$, $10^7M_{\odot}$ and $10^8M_{\odot}$ respectively.}
\label{figure9}
\end{figure}

Fig.~\ref{figure9} shows the distribution of the closest separation as a function of time. The blue, green and red curves correspond to the results for SMBH masses $M=10^6M_{\odot}$, $10^7M_{\odot}$ and $10^8M_{\odot}$, respectively. We find that the cumulative closest separation distributions are very similar for the different SMBH mass, but the closest separation distribution for a more massive SMBH is more flat. 

\begin{figure}
\centering
\includegraphics[width=\columnwidth]{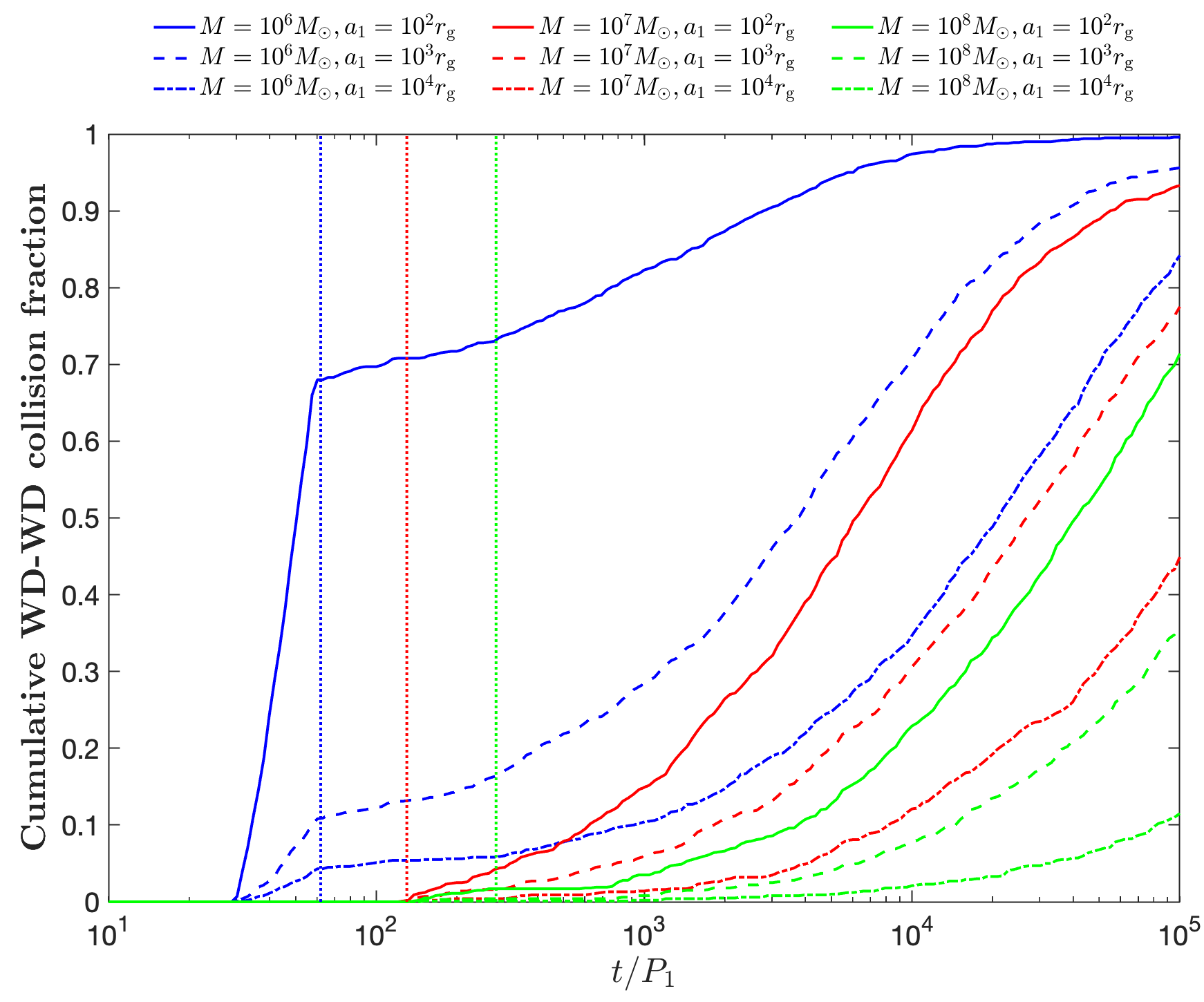}
\vspace*{-4mm}
\caption{Cumulative WD-WD collision fraction as a function of time. The blue, red and green curves correspond to the results for $M=10^6,10^7,10^8M_{\odot}$ respectively. The solid curves represent $a_1 = 10^2r_{\rm g}$, the dashed curves represent $a_1 = 10^3r_{\rm g}$ and the dashed-dotted curves $a_1 = 10^4r_{\rm g}$. The dotted lines represents $t=62P_1$.}
\label{figure10}
\end{figure}

More massive SMBHs have larger $r_{\rm g}$. This indicates that, at $a_1 = 10^2r_{\rm g}$, the ratio between WD-WD collision separation and the orbital radius (i.e. $r_{\rm c}/a_1$) will decrease as SMBH mass increases. Fig.~\ref{figure10} shows the cumulative distribution of WD-WD collision at different initial radii for different SMBH masses. We find that, as expected, WD-WD collisions decrease as SMBH mass increases. Unlike $M = 10^6M_{\odot}$, there are almost no WD-WD collisions during the first two encounters  for $M = 10^7M_{\odot}$ and $10^8M_{\odot}$. At $a_1=10^4r_{\rm g}$, only $\sim 10$ per cent of samples have WD-WD collisions for $M=10^8M_{\odot}$. If we consider migration and $10^5P_1\approx10{\rm Myr}$ for $M=10^8M_{\odot}$ and $a_1=10^4r_{\rm g}$, the WD-WD collision fraction will be lower. As a result, WD-WD collisions in our model should occur preferentially in AGNs with less massive SMBHs. 

\subsection{WD-WD Collision rate}
\label{section3.4}
The number of BHs around the SMBH is about $\sim 1-4\times 10^4$ \citep{2018MNRAS.478.4030G}. According to the stellar IMF \citep{2001MNRAS.322..231K}, the number of WDs are ten times larger than that of BHs. As a result, $\sim 2\times 10^5$ WDs should exsit around the central SMBH. We assume that the distribution of the stellar mass is cuspy and  the initial number density of WDs is given as follow \citep{2021ApJ...908..194T},
\begin{equation}
\label{eq:8}
\frac{dN_{_{\rm WD}}(r)}{dr} \propto r^{-0.5},
\end{equation}
where $N_{_{\rm WD}}(r)$ is the total number of WDs within distance $r$ from the SMBH. 

Due to type I migration, the orbital separation between WDs will change with time. As their orbital separation reaches $\sim 3R_{\rm H}$, they will form a "SMBH+WD+WD" systems. For such a system, the WD-WD collision rate can be written as 
\begin{equation}
\label{eq:9}
\mathcal{R} = n_{_{\rm GN}} \times f_{_{\rm AGN}}\times \frac{N_{\rm WD}\times f_{{\rm d}} \times f_{\rm 3b}\times f_{{\rm c}}}{\tau_{_{\rm AGN}}},
\end{equation} 
where $n_{_{\rm GN}}$ is the number density of galactic nuclei in the Universe, $f_{_{\rm AGN}}$ is the fraction of galactic nuclei that have active AGNs, $\tau_{_{\rm AGN}}$ is the lifetime of AGNs, $N_{\rm WD}$ is the total number of WDs within distance $10^6r_{\rm g}$ from the central SMBH, $f_{{\rm d}}$ is the fraction of WDs that end up in AGN disc, $f_{\rm 3b}$ is the probability of forming a three-body system, and $f_{{\rm c}}$ is the fraction of WD-WD collisions in such a three-body system. Our finding indicates that WD-WD collisions occur in nearly all samples at the inner region of the discs ($r < 10^4r_{\rm g}$) if we extend the simulation time to the AGN lifetime, but at the outer region of the disc ($a_1 > 10^4r_{\rm g}$) the WD-WD collision fraction will decrease quickly, especially for SMBHs with large mass. We therefore assume that, as the WDs migrate to $r < 10^4r_{\rm g}$, the WD-WD collision fraction $f_{\rm c} = 1$, for other case $f_{\rm c} = 0$. We also assume that, for all AGNs, half of the WDs in AGN discs can migrate to $r < 10^4r_{\rm g}$ and form restricted three-body systems, which gives $f_{\rm 3b} = 0.25$. Thus, we can get an approximately WD-WD collision rate 
\begin{equation}
\label{eq:10}
\begin{aligned}
\mathcal{R} = \: &300{\rm Gpc^{-3} yr^{-1}}\frac{n_{_{\rm WD}}}{0.006{\rm Mpc^{-3}}} \frac{N_{\rm WD}}{2\times 10^5} \frac{f_{_{\rm AGN}}}{0.1} \frac{f_{\rm d}}{0.1} \\
                       &\times \frac{f_{\rm 3b}}{0.25} \frac{f_{\rm c}}{1} \left(\frac{\tau}{10 {\rm Myr}}\right)^{-1}.
\end{aligned}
\end{equation} 
If all of those WD-WD collisions can produce type I SNe, we have an overall event rate 300 ${\rm Gpc^{-3}yr^{-1}}$ in AGN discs. The observed rate of type Ia SNe in the local universe is $2.5\pm 0.5\times 10^4 {\rm Gpc^{-3}yr^{-1}}$ \citep{2011MNRAS.412.1473L, 2015A&A...584A..62C}. Therefore, WD-WD collisions only constitute roughly 1 per cent of the observed rate of SN Ia.

There are some caveats and conditions that can influence our estimation of the event rate. First, there are uncertainties in the properties of the AGNs and the distributions of WDs. The number of WDs in AGN discs likely depends on the mass density of discs and the distribution of WDs. For discs with large surface density $\Sigma$, the WDs more likely align with the discs \citep{2019ApJ...876..122Y}. In addition to the alignment, the migration of WDs is also affected by the surface density of the AGN discs and the mass of the central SMBHs. Around a  more massive SMBH, such as $10^8M_{\odot}$, the time-scale of the migration gets longer by a factor $10^2$ (because $r_{\rm g} = GM/c^2$) compared with the case of $10^6M_{\odot}$, in which a WD starts at $R = 100r_{\rm g}$. In summary, migration-driven collisions more likely occur around less massive SMBHs. In the most semi-realistic disc models \citep{2003MNRAS.341..501S, 2005ApJ...630..167T}, the surface density $\Sigma$ likely drops off at $\sim 10^3r_{\rm g}$ and $h/R$ increases, leading to an increase in 
the time-scale of migration of WDs, but substantial changes of orbital radius can occur for WDs in the inner disc ($<10^4r_{\rm g}$) over the lifetime of AGNs  ($\sim 10\rm {Myr}$) \citep{2012MNRAS.425..460M}. At large radius ($> 10^4r_{\rm g}$), the migration of WDs into the inner disc might be less likely.

Second, in addition to migration, three additional factors are likely to be important in driving collisions of WDs. (i) Some WD binaries may exsit that are formed through binary stellar evolution or dynamical process. Those WD binaries are likely to be dynamically hard in the nucleus, otherwise they might be ionized via dynamical encounters. Those hard WD binaries could be driven to merge via three-body scattering and due to the gas effects within the disc. (ii) If the turbulence scales of the discs are large enough (e.g. $>2\sqrt{3}R_{\rm H}$), WDs might not encounter each other due to the migration, but they might collide via random encounters. (iii) The dynamics might drive random encounters. WDs could collide randomly via direct collisions before they align with AGN discs. Even after they align with AGN discs, they might undergo random collisions during the damping of their eccentricity.

Third, other compact objects (BHs and NSs) and stars in AGN discs, neglected in this study, might influence WD-WD collisions. Since the migration rate for BHs and NSs in the AGN disc is fast, BHs and NSs  will easily encounter WDs via fast migration and can tidally disrupt them during their close counters \citep{2012MNRAS.419..827M}, which may decrease the population of WDs in the disc. In addition to encounters with compact objects, WDs can also encounter stars in the disc, and this might lead to the explosions of both stars.

\section{Conclusion and Discussions}
\label{section4}

In this work, we study the closest separation between two WDs and WD-WD collisions via close encounters in AGN discs. We perform a set of N-body simulations with different initial parameters, including the initial orbital separation, the relative orbital inclination of the WDs, and different masses of SMBHs. According to our analysis and simulations, the WDs in AGN discs cannot form binary WD systems via the emission of gravitational wave. Our finding are as follows:

In Run1-Run7, the initial orbital separation changes from $0.8R_{\rm H}$ to $4.0R_{\rm H}$. It is found that close encounters between  WD-WD only occur at $p \lesssim 2\sqrt{3} R_{\rm H}$. For $1.1R_{\rm H} \lesssim p \lesssim 2\sqrt{3}R_{\rm H}$, the closest separation between two WDs is concentrated between $10^{-8}a_1$ and $10^{-6} a_1$. As $p \lesssim 1.1R_{\rm H}$, most WDs enter horseshoe or tadpole orbits, which leads to fewer WD-WD close encounters (see Figs~\ref{figure4} and~\ref{figure5}). In the case of $p=3.0R_{\rm H}$, the WD-WD collision fraction decreases as the initial orbit radius $a_1$ (Fig.~\ref{figure3}) increases. In the inner region of the disc ($r\lesssim10^3r_{\rm g}$), considerable WD-WD collisions occur at $t < 62P_1$(Fig.~\ref{figure6}). We note that WD-WD collisions in our systems can occur at $a_1 < 10^5r_{\rm g}$ for most of the samples within $10^5P_1$.

Taking the relative orbital inclinations into consideration, the peak of the closest separation distribution increase as the relative inclinations increase (Fig.~\ref{figure7}). For small relative inclinations (i.e. $\Delta{i} \lesssim 10^{-2}R_{\rm H}/a_1$), the fraction of WD-WD collisions is almost the same as that in the case of coplanar orbits. However, for large relative inclinations (i.e. $\Delta{i} > 10^{-2}R_{\rm H}/a_1$), the WD-WD collision fraction decreases quickly as the relative inclination increases and WD-WD collisions become rare at $a_1 \leq 10^4r_{\rm g}$ (Fig.~\ref{figure8}).

It is clearly shown in Fig.~\ref{figure9} that the closest separation distribution is very similar in case of different SMBH mass. For more massive SMBHs, however, the ratio between collision separation and orbital radius is smaller, which leads to fewer events of WD-WD collisions for more massive SMBHs. For $M=10^8M_{\odot}$, there are nearly no WD-WD collisions at $a_1=10^4r_{\rm g}$ (Fig.~\ref{figure10}). That means WD-WD collisions due to migration might happen preferentially in AGNs around less massive SMBHs.

According to our rough estimation, the event rate of WD-WD collisions is around 300${\rm Gpc^{-3}yr^{-1}}$. At most, such WD-WD collisions can contribute roughly 
1 per cent of the overall event rate of Type Ia SNe. These SN explosions in AGN discs can generate strong shocks, which can lead to unique observations. In our following work, we will discuss the observational characteristics of SNe Ia in AGN discs resulting from WD-WD collisions.

A few uncertainties exsit in our N-body simulations. The most important one is the effects of disc gas. During evolution, frictional disc forces can affect the dynamical evolution of WDs. \citep{2022arXiv220305584L} have studied the effect of the disc forces with simple prescriptions; their results suggest that disc forces have little effect on the very close encounters. In addition to disc forces, WDs in AGN disc usually accrete gas. Nova reactions may exsit on the surface of a WD as it accretes hydrogen from the AGN disc. The feedback of the nova could puff the gas off from the WD and influence the migration of the WD. Once the mass of WDs grow to the Chandrasekhar limit, WDs produce 
SN Ia. However, it is difficult for WDs to grow to the Chandrasekhar limit via accretion, because WDs are spun up more efficiently to reach the shedding limit before the Chandrasekhar limit \citep{2021ApJ...923..173P}. Besides the growth of their mass, mini-discs around the WDs might help damp their orbital energy as they encounter each other and facilitate the formation of binary WDs  \citep{2023ApJ...944L..42L}. Furthermore, accretion also affects the mass ratio of the two WDs. However, the masses ratio affects the close encounter rate by a factor of $\sim 2$ \citep{2022arXiv220305584L}. Thus, our results are roughly in keeping in systems with different mass ratio of WD-WD. 

In this work, we study  WD-WD collisions via close encounter in AGN discs. However, WD-BH, WD-NS and star-BH collisions also likely happen in AGN discs. In our following work, we will study close encounters of WD-BH, WD-NS and star-BH, which may lead to 'micro' Tidal Disruption Events (TDEs) and the subsequent light variation of AGNs \citep{2021ApJ...915...10P,2021MNRAS.507..156G}. 

\section*{Acknowledgements}
We would like to thank the referee for her/his critical comments which are very helpful for the improvement of this paper. 
YFY was supported by the National SKA Program of China
No. 2020SKA0120300, and the National Natural Science Foundation of
China (Grant No. 11725312).  LCH was supported by the National Science
Foundation of China (11721303, 11991052, 12011540375) and the
China Manned Space Project (CMS-CSST-2021-A04, CMS-CSST-
2021-A06). JMW acknowledges support from the National Science
Foundation of China (NSFC-11833008 and -11991054) and from
the National Key R\&D Program of China (2016YFA0400701). The
numerical calculations in this paper have been done on the 
supercomputing system in the Supercomputing Center of University of
Science and Technology of China.
%%%%%%%%%%%%%%%%%%%%%%%%%%%%%%%%%%%%%%%%%%%%%%%%%%
\section*{Data Availability}
The data underlying this article will be shared on reasonable request to the corresponding author (YFY).
%The inclusion of a Data Availability Statement is a requirement for articles published in MNRAS. Data Availability Statements provide a standardised format for readers to understand the availability of data underlying the research results described in the article. The statement may refer to original data generated in the course of the study or to third-party data analysed in the article. The statement should describe and provide means of access, where possible, by linking to the data or providing the required accession numbers for the relevant databases or DOIs.



\bibliographystyle{mnras}
\bibliography{WDcollision} % if your bibtex file is called example.bib

%\begin{thebibliography}{99}
%\end{thebibliography}

%%%%%%%%%%%%%%%%%%%%%%%%%%%%%%%%%%%%%%%%%%%%%%%%%%

%%%%%%%%%%%%%%%%% APPENDICES %%%%%%%%%%%%%%%%%%%%%

%%%%%%%%%%%%%%%%%%%%%%%%%%%%%%%%%%%%%%%%%%%%%%%%%%

% Don't change these lines
\bsp	% typesetting comment
\label{lastpage}
\end{document}